\documentclass[a4paper,11pt]{article}
\usepackage{jheppub}
\usepackage{tikz-cd}
\usetikzlibrary{calc, shapes.misc}
\usepackage{longtable}
\usepackage{orcidlink}
\usepackage{subcaption}
\tikzset{
			base/.style={baseline=-0.6ex, scale=1},
			dot/.style={circle, fill=black, inner sep=1.5pt, outer sep=0pt},
			cross/.style={font=\bfseries\scriptsize, inner sep=0pt},
			tube/.style={draw, thick, rounded corners=4.5pt, line width=0.8pt},
			chain/.style={thick}
		}
\usepackage{array}
\newcolumntype{C}[1]{>{\centering\arraybackslash}m{#1}}

\title{Cluster Bootstrap for Cosmological Correlators}

\author[a, \orcidlink{0000-0003-2613-718X}]{\small Shruti Paranjape,}
\emailAdd{shruti\_paranjape@brown.edu}

\author[a, \orcidlink{0000-0003-2569-1234}]{\small Marcos Skowronek,}
\emailAdd{marcos\_skowronek\_santos@brown.edu}

\author[a, b, \orcidlink{0009-0005-6084-2466}]{\small Marcus Spradlin,}
\emailAdd{marcus\_spradlin@brown.edu}

\author[a, \orcidlink{0009-0008-2506-3207}]{\small Anastasia Volovich,}
\emailAdd{anastasia\_volovich@brown.edu}

\author[a, \orcidlink{0000-0003-3781-6153}]{\small and He-Chen Weng}
\emailAdd{he-chen\_weng@brown.edu}

\affiliation[a]{\footnotesize Department of Physics,
    Brown University,
    Providence,
    RI 02912,
    USA
}

\affiliation[b]{\footnotesize Brown Center for Theoretical Physics and Innovation,
    Brown University,
    Providence,
    RI 02912,
    USA
}

\abstract{We show that cosmological wavefunction coefficients associated with $n$-site chain and loop graphs for a cubic scalar theory in de Sitter spacetime have symbol alphabets given by subsets of $A_{2n{-}2}$ and $B_{2n{-}1}$ cluster variables, respectively, and satisfy the associated cluster adjacency properties. The key step in proving this is identifying a precise connection
between graph ``tubings'' that appear in the kinematic flow equation and polygon ``triangulations'' that encode
the combinatorics of cluster compatibility.
Our results imply that cosmological wavefunction coefficients in a general power-law FRW cosmology satisfy cluster
adjacency to all orders in the $\epsilon$ expansion around the de Sitter limit.
We use this information as bootstrap input to show that de Sitter symbols for $n \le 4$ are uniquely determined by simple physical constraints.
}

\begin{document}
\maketitle
\flushbottom

\section{Introduction}

    Cluster algebras have played a surprising role in describing many mathematical properties of the singularity structure of multi-loop amplitudes in $\mathcal{N}$=4 supersymmetric Yang-Mills (SYM).  For 6- and 7-particle amplitudes, all symbol letters~\cite{Goncharov:2010jf} of amplitudes are given~\cite{Golden:2013xva} by certain cluster variables~\cite{Fomin2001ClusterAI, gekhtman2002cluster, Fomin:2016caz}, and the mathematical notion of cluster compatibility describes which letters may appear adjacent to one another in a symbol~\cite{Drummond:2017ssj,Golden:2019kks}.  The hypothesis that these properties remain true to all loop orders serves as the foundation of a bootstrap program (see~\cite{Caron-Huot:2020bkp} for review) that has made possible the computation of the symbol for 8-loop 6-particle~\cite{Dixon:2023kop} and 5-loop 7-particle~\cite{He:2025tyv} amplitudes. The connection between symbol alphabets and cluster algebras has recently been extended beyond the realm of SYM theory, including to integrals relevant for QCD \cite{Chicherin:2020umh, Aliaj:2024zgp, Pokraka:2025ali, Bossinger:2025rhf}.
	
	In recent years, many techniques to calculate scattering amplitudes have been adapted to the study of scalar wavefunctions in cosmology, and many new mathematical structures have been uncovered. These include the singularity structure of their integrands via positive geometries~\cite{Arkani-Hamed:2017fdk, Benincasa:2020aoj, Benincasa:2024leu, Arkani-Hamed:2024jbp, Glew:2025otn, Arundine:2026fbr}, hidden zeros of integrands~\cite{De:2025bmf}, differential equations for integrals~\cite{De:2023xue, Arkani-Hamed:2023kig, De:2024zic, Capuano:2025ehm}, and recursion relations satisfied by integrals~\cite{Hillman:2019wgh, He:2024olr}. The study of such cosmological wavefunction coefficients is crucial to the construction of cosmological correlators which in turn make concrete predictions for observations of early universe physics; see~\cite{Benincasa:2022gtd} for a review.
	
	In particular, (de Sitter) FRW wavefunction coefficients are given by (un)twisted integrals whose integrands are shifted flat space wavefunction coefficients~\cite{Arkani-Hamed:2017fdk}. Here, it is important to ask what the function space of these integrals is, and analogous to the case of amplitudes in SYM theory, the answer lies in the structure of the corresponding symbols. The symbol alphabet of graphs that contribute to the wavefunction coefficient has been studied previously in some generality~\cite{Hillman:2019wgh}. For graphs with chain topology, it was shown recently that the symbol alphabet is given by (a subset of) the cluster variables of $A$-type cluster algebras~\cite{Mazloumi:2025pmx, Capuano:2025myy}.
	
	In this paper, we first extend this result by showing that the symbol alphabet of the wavefunction coefficient associated with the $n$-gon (also called the $n$-site loop) graph is given by (a subset of) the cluster variables of the $B_{2{n}-1}$ cluster algebra. We then use the rules of kinematic flow~\cite{Arkani-Hamed:2023kig} that describe the differential equations satisfied by cosmological wavefunction coefficients to prove that, for both chain and loop graphs, the corresponding symbols satisfy cluster adjacency with respect to the corresponding $A$- and $B$-type algebras.  This is achieved by identifying a precise correspondence between complete tubings and clusters. We also initiate a cluster bootstrap\footnote{A different type of cosmological bootstrap was described in~\cite{Arkani-Hamed:2018kmz}.} that starts from these symbol alphabets and adjacency conditions, and shows that the symbols of the corresponding de Sitter wavefunction coefficients are (up to an unfixed overall normalization) uniquely determined for $n=2,3,4$ by imposing certain elementary physical properties including a first-entry condition and vanishing in soft limits. Interestingly, we find that for chain graphs, cluster adjacency emerges as a consequence of the other conditions and is a redundant input to the bootstrap, whereas for loop graphs, it is essential to assume cluster adjacency in order for the bootstrap to land on a unique symbol.  
	
	The paper is organized as follows. In Section~\ref{sec:rev} we briefly review the kinematic flow equations for cosmological wavefunction coefficients and a few necessary facts about cluster variables and compatibility.  In Sections~\ref{sec:chain graphs} and~\ref{sec:gon graphs} we show that the de Sitter wavefunction coefficients corresponding to $n$-site chain and loop graphs have symbol alphabets given by subsets of cluster variables associated with $A_{2n{-}2}$ and $B_{2n{-}1}$, respectively, and we show that it follows from the kinematic flow equations that their symbols satisfy cluster adjacency with respect to these algebras.  We also discuss the bootstrap for these graph topologies.  We end in Section~\ref{sec:outlook} with a discussion.

\smallskip
\noindent
{\bf Note added}: Upon the completion of this work we learned of the paper~\cite{Capuano:2026pgq} which describes a generalized notion of cluster adjacency satisfied by the $n$-chain graphs.

\section{Review}
\label{sec:rev}

In this section, we review two ingredients that are key to our work: first, the rules of kinematic flow introduced in \cite{Arkani-Hamed:2023kig} that describe the differential equations satisfied by cosmological wavefunction coefficients in de Sitter space, and second, some basics of cluster algebras including cluster variables and compatibility~\cite{Fomin2001ClusterAI,Fomin:2002nsg,Fomin:2016caz, Fomin2017}.

\subsection{Symbols for dS wavefunction coefficients}
\label{sec:review_cosmo}

De Sitter wavefunction coefficients are elements of a vector $\vec{I}$ of functions satisfying the differential equation\footnote{Here we have taken the $\epsilon \to 0$ limit of the differential equation that applies for a general power-law FRW cosmology to specialize to our immediate case of interest, which is de Sitter. In this case, the equation truncates~\cite{Arkani-Hamed:2023kig}; we comment on the application of our results to more general cases in Section~\ref{sec:outlook}.}
\begin{equation}
\label{eq:dAI}
    d \vec{I} = A \vec{I}
\end{equation}
where $d$ stands for the total differential with respect to all kinematic variables.
The wavefunction coefficient associated with a Feynman graph $\mathcal{G}$ depends on two types of kinematic variables  associated with the graph: one variable $X_v$ attached to each vertex and one variable $Y_e$
attached to each internal edge.  These $X$'s and $Y$'s are simple sums of the ``energies''
$|\vec{ k}_i|$ of the particles involved in the process; see \cite{Arkani-Hamed:2023kig}.

The matrix $A$ appearing in~(\ref{eq:dAI}) has the form
\begin{equation}
	A=\sum_i \alpha_i d\log\Phi_i(X_j,Y_k)
\end{equation}
where the $\alpha_i$ are constant nilpotent matrices.
This form ensures that the solutions to~(\ref{eq:dAI}) are multiple polylogarithm functions with
symbol letters given by the set of $\Phi_i$.

The symbol alphabet can be read off from the graph as explained in \cite{Arkani-Hamed:2023kig}:
it is the set of region variables corresponding to all possible tubes of the marked graph of $\mathcal{G}$. The markings are placed on all internal edges of $\mathcal{G}$ and indicate certain sign flips in the internal energies.
In this paper, we only consider $n$-chains and $n$-gons,
so we can simply label each tube by the first
and last vertex that it encloses.  Given a tube enclosing vertices $i$ through ${j}$ (inclusive),
the corresponding symbol letter is
\begin{equation}
    \sum_{k=i}^{j} X_{k} \pm Y_{i{-}1} \pm Y_{j}
\end{equation}
where the sign of $Y_{i{-}1}$ is determined to be $-$ or $+$ respectively depending on whether or not the marking $i-1$ is enclosed in the tube. Similarly the sign of $Y_{j}$ is determined by the enclosure of marking $j$\footnote{Here we set $Y_0=Y_n=0$ for the $n$-site chain graph and $Y_0=Y_n$ for the $n$-gon.}. For example, the symbol letters associated
to the following tubes are
\begin{equation}
	\begin{aligned}
		\tikzset{
			base/.style={baseline=-0.6ex, scale=1},
			dot/.style={circle, fill=black, inner sep=1.5pt, outer sep=0pt},
			cross/.style={font=\bfseries\scriptsize, inner sep=0pt},
			tube/.style={draw, thick, rounded corners=4.5pt, line width=0.8pt},
			chain/.style={thick}
		}
		\newcommand{\drawchain}{
			\draw[chain] (0,0) -- (3.2,0);
			\node[dot] at (0,0) {};    
            \node at (0,-0.4) {\small $X_1$};
			\node[cross] at (0.8,0) {$\times$}; 
            \node at (0.8,0.4) {\small $Y_1$};
			\node[dot] at (1.6,0) {};  
            \node at (1.6,-0.4) {\small $X_2$};
			\node[cross] at (2.4,0) {$\times$}; 
            \node at (2.4,0.4) {\small $Y_2$};
			\node[dot] at (3.2,0) {};  
            \node at (3.2,-0.4) {\small $X_3$};
		}
		\begin{tikzpicture}[base]
			\drawchain
			\draw[tube] (0.6, -0.16) rectangle (1.8, 0.16);
		\end{tikzpicture}
		&=  X_2 - Y_1 + Y_2\,, \\[1em]
		\tikzset{
			base/.style={baseline=-0.6ex, scale=1},
			dot/.style={circle, fill=black, inner sep=1.5pt, outer sep=0pt},
			cross/.style={font=\bfseries\scriptsize, inner sep=0pt},
			tube/.style={draw, thick, rounded corners=4.5pt, line width=0.8pt},
			chain/.style={thick}
		}
		\newcommand{\drawchain}{
			\draw[chain] (0,0) -- (3.2,0);
			\node[dot] at (0,0) {};    
            \node at (0,-0.4) {\small $X_1$};
			\node[cross] at (0.8,0) {$\times$}; 
            \node at (0.8,0.4) {\small $Y_1$};
			\node[dot] at (1.6,0) {};  
            \node at (1.6,-0.4) {\small $X_2$};
			\node[cross] at (2.4,0) {$\times$}; 
            \node at (2.4,0.4) {\small $Y_2$};
			\node[dot] at (3.2,0) {};  
            \node at (3.2,-0.4) {\small $X_3$};
		}
		\begin{tikzpicture}[base]
			\drawchain
			\draw[tube] (1.4, -0.16) rectangle (3.4, 0.16);
		\end{tikzpicture}
		&=  X_2 + X_3 + Y_1\,.
	\end{aligned}
\end{equation}

\begin{figure}[t]
	\centering
	\vspace{0.5cm}
	\resizebox{0.7\linewidth}{!}{%
		\begin{tikzpicture}[
			x=1.4cm, 
			y=1.0cm, 
			dot/.style={circle, fill=black, inner sep=2pt, outer sep=0pt},
			cross/.style={font=\bfseries\large},
			txt/.style={font=\normalsize}, 
			tube/.style args={#1/#2}{
				draw=#2, thick, line width=1pt, 
				rounded corners=#1
			}
			]
			
			\newcommand{\drawchain}[2]{
				\begin{scope}[shift={(#1,0)}]
					\draw[thick] (0,0) -- (4,0);
					
					\foreach \i [count=\x from 0] in {1,...,5} {
						\node[dot] (X\i_#2) at (\x, 0) {};
						\node[txt] at (\x, -0.8) {$X_{\i}$};
					}
					
					\foreach \i [count=\x from 0] in {1,...,4} {
						\coordinate (Y\i_pos) at (\x + 0.5, 0);
						\node[cross] (Y\i_#2) at (Y\i_pos) {$\times$};
						\node[txt] at (\x + 0.5, 0.8) {$Y_{\i}$};
					}
				\end{scope}
			}
			
			
			\drawchain{0}{L}
			
			\draw[tube=0.35cm/black] 
			($(X1_L)+(-0.25, 0.35)$) rectangle ($(X3_L)+(0.22, -0.35)$);
			
			\draw[tube=0.22cm/black] 
			($(X2_L)+(-0.25, 0.22)$) rectangle ($(X3_L)+(0.15, -0.22)$);

			
			\drawchain{5}{R}
			
			\draw[tube=0.3cm/black] 
			($(X1_R)+(-0.25, 0.3)$) rectangle ($(X3_R)+(0.22, -0.3)$);
			
			\draw[tube=0.22cm/black] 
			($(X3_R)+(-0.25, 0.22)$) rectangle ($(X4_R)+(0.15, -0.22)$);
			
		\end{tikzpicture}%
	}
	\caption{Examples of a valid (left) and forbidden (right) pair of tubings corresponding to adjacent entries in the symbol for the 5-site chain graph.}
	\label{fig: examples tubings chain}
\end{figure}
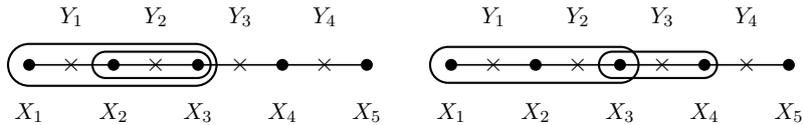

The paper~\cite{Arkani-Hamed:2023kig} also explained how to read off the differential equation~(\ref{eq:dAI}) from the graph.
The differential equation is written in a basis of integrals indexed by maximal sets of non-overlapping tubings (called complete tubings) of $\mathcal{G}$.  For each complete tubing $t$, the differential
of the associated basis element $I_t$ is determined
by the following ``kinematic flow'' rules:
\begin{enumerate}
	\item Choose a tube in the complete tubing $t$ to activate, and draw its descendant tubings via growth, merger, and absorption, while keeping track of the activated tube.
    \item The set of all activated tubes and their descendants is then referred to as the kinematic flow of $t$. Each node $i$ in this flow is associated with a letter $\Phi_i$ and an integral $I_i$.
	\item Then row $t$ of the differential equation~(\ref{eq:dAI}) is
	\begin{align}\label{eq:kinflow}
		dI_t = \sum_i (I_i - \sum_j I_{ij}) d\log \Phi_i\,, 
	\end{align}
	where we sum over all nodes $i$ of the kinematic flow of $t$, $I_i$ is the basis integral of node $i$, $I_{ij}$ are the functions associated with the immediate descendants of node $i$ and $\Phi_i$ is the symbol letter corresponding to node $i$.
\end{enumerate}
We refer the reader to~\cite{Arkani-Hamed:2023kig} for more details, including several examples. Here, we only need the following observation. None of the letters ${\Phi_i}$ appearing in \eqref{eq:kinflow} correspond to tubes that intersect the tubing associated with $I_i$ or its descendants $I_{ij}$. When applied recursively, we see that no single term in the symbol can contain adjacent letters that correspond to tubings that intersect (see Figure~\ref{fig: examples tubings chain} for an example).

\subsection{Cluster variables and compatibility}
\label{sec:rev_cluster}

Cluster algebras are generated by sets (called clusters) of variables related to each other by an operation called mutation; for an introduction and review see~\cite{Fomin2001ClusterAI,Fomin:2002nsg,Fomin:2016caz, Fomin2017}.

For the $A_n$ cluster algebra each cluster corresponds to a triangulation of an $(n{+}3)$-gon,
with cluster variables associated with each of the $n$ internal chords and $n+3$ boundary chords.
An explicit realization of this algebra is provided by identifying the variable associated with
the chord connecting vertices $i$ and $j$ with the Pl\"ucker coordinate $\langle i\,j\rangle$ on
the Grassmannian Gr$(2,n+3)$. Parameterizing an element of this Grassmannian by
\begin{equation}\label{eq: A type matrix param}
    \begin{pmatrix}
        1 & 1 & 1 & \dots & 1 & 1 & 0 \\
        0 & z_1 & z_2 & \dots & z_n & 1 & 1
    \end{pmatrix}\,,
\end{equation}
the Pl\"ucker coordinates are the $2 \times 2$ minors of this matrix:
\begin{align}\label{eq: A_n cluster variables}
\begin{aligned}
\Delta_{i,j} &= z_{j - 1} - z_{i-1}\,, \qquad &1 \le i < j \le n+2\,,\\
\Delta_{i,n+3} &= 1\,, \qquad &1 \le i \le n+2\,,
\end{aligned}
\end{align}
with the understanding that $z_0=0$ and $z_{n+1}=1$.
We will use the notation $\Delta_{i,j}$ both for the chord and the corresponding Pl\"ucker coordinate.
Two cluster variables are said to be \emph{compatible} if they appear in a common cluster.
The variables associated with two chords are compatible if and only if the chords do not cross.

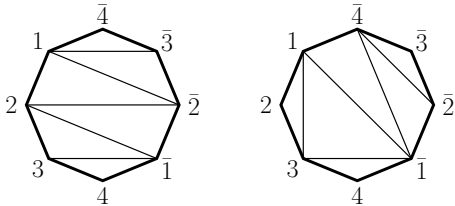
\begin{figure}[htbp]
\centering
\resizebox{0.4\linewidth}{!}{%
\begin{tikzpicture}[
    scale=1, 
    every node/.style={font=\LARGE}, 
    thick 
]
\newcommand{\drawoctagon}[2]{
    \begin{scope}[#1]
        \def\R{1.8} 
        
        \foreach \i/\lab/\ang in {
            1/1/135, 2/2/180, 3/3/225, 4/4/270, 
            5/\bar{1}/315, 6/\bar{2}/0, 7/\bar{3}/45, 8/\bar{4}/90
        } {
            \coordinate (v\i) at (\ang:\R);
            \node at (\ang:\R+0.35) {$\lab$};
        }
        
        \draw[line width=2pt] (v1) -- (v2) -- (v3) -- (v4) -- (v5) -- (v6) -- (v7) -- (v8) -- cycle;
        
        #2
    \end{scope}
}
\drawoctagon{}{
    \draw (v2) -- (v6);
    \draw (v1) -- (v6); 
    \draw (v1) -- (v7); 
    \draw (v3) -- (v5); 
    \draw (v2) -- (v5); 
}
\drawoctagon{shift={(6,0)}}{
    \draw (v1) -- (v3);
    \draw (v1) -- (v5); 
    \draw (v3) -- (v5); 
    \draw (v6) -- (v8); 
    \draw (v5) -- (v8); 
}
\end{tikzpicture}%
}
\caption{Left: example of a symmetric triangulation of the octagon, corresponding to a cluster in $B_3$. Right: example of an invalid triangulation.}
\label{fig: triangulations B-type}
\end{figure}

For the $B_n$ cluster algebra, each cluster corresponds to a triangulation of a $(2n{+}2)$-gon that is symmetric under a $180^\circ$ rotation (see Figure \ref{fig: triangulations B-type}). Again, the chords in each triangulation, both internal and boundary, correspond to the variables in a cluster. It is convenient to label the boundary vertices of the $(2n{+}2)$-gon by $\{1,2,\dots n+1, \overline{1}, \overline{2},\dots \overline{n+1}\}$. Symmetry implies that the set of chords of a valid triangulation must be invariant under the action $a \leftrightarrow \overline{a}$. Just like for the $A_n$ cluster algebra, each chord is also uniquely mapped to a cluster variable:
\begin{align}
    \Delta_{a,b} = \Delta_{\overline{a},\overline{b}} =  z_{b} - z_{a}\,,\quad \Delta_{a,\overline{b}}= \Delta_{\overline{a},b} = 1-(z_{b}-z_{a})\,,\quad 1\leq a\leq b\leq n+1\,.
\end{align}
Again, cluster compatibility for $B_n$ cluster algebras has a simple geometric interpretation: a pair of chords is cluster compatible if and only if the two chords and their $a \leftrightarrow \overline{a}$ conjugates do not cross.

\section{Chain Graphs}
\label{sec:chain graphs}

In this section we explore the cluster algebra structure corresponding to chain graphs. In Section~\ref{sec:chainmap} we 
 demonstrate a map between the symbol alphabet of the wavefunction coefficient associated with the $n$-site chain, which
is a weight-$n$ multiple polylogarithm function, and the cluster variables of the $A_{2n{-}2}$ cluster algebra. This map differs from the one recently discussed in~\cite{Mazloumi:2025pmx} and is the same as the one in \cite{Capuano:2025myy}. 
In Section \ref{sec: cluster adj chain}
we demonstrate that the symbols satisfy cluster adjacency with respect to this algebra.
Then, after a few explicit examples in Sections~\ref{sec:chain2} and~\ref{sec:chain3}, we discuss
the cluster bootstrap for computing these symbols in Section~\ref{sec:chainboot}.

\subsection{Symbol alphabets from \texorpdfstring{$A$}{A}-type cluster algebras}
\label{sec:chainmap}

The symbol alphabet of the $n$-site chain graph can be read off the tubings of the marked graph. These are 
\begin{align}
\label{eq:chainsym}
    &X_{j,k}^{\pm\pm}\equiv\sum_{i=j}^k X_i \pm Y_{j-1}\pm Y_k\,,\quad1\le j\le k\le n\,,\quad 
    {\rm and} \quad  X\equiv\sum_{i=1}^n X_i\,,
\end{align}
where we define $Y_0=Y_n=0$.
This comprises a list of $2n(n{-}1){+}1$ symbol letters that can be assembled
into an alphabet of $2n(n{-}1)$ dimensionless ratios.

It was recently shown in~\cite{Capuano:2025myy,Mazloumi:2025pmx} that the letters of this symbol alphabet can be identified with a subset of those associated with the $A_{2n{-}2}$ algebra.
Specifically, the cluster variables of the latter assemble naturally into $(2n{+}1)(n{-}1)$ ratios that include
the $2n(n-1)$-letter symbol alphabet of the $n$-site chain together with $n{-}1$ additional spurious letters that do not appear in the symbols of chain graph wavefunction coefficients.

The map provided in \cite{Mazloumi:2025pmx} is sufficient for establishing that symbol letters are cluster variables, but does not make cluster adjacency manifest.  Therefore we will use the map of~\cite{Capuano:2025myy}, which allows us to straightforwardly manifest the fact that the symbol entries obey the cluster adjacency conditions. Specifically we use the parameterization~(\ref{eq: A type matrix param}) with entries $z_i$ given in terms of the
kinematic data of the $n$-chain graph by:
\begin{equation}\label{eq:cluster map chain}
    \boxed{z_{2i-1} = \frac{1}{X}{\left(\sum_{k=1}^i X_k + Y_i\right)}\,,\quad z_{2i} = \frac{1}{X}{\left(\sum_{k=1}^i X_k - Y_i \right)}\,, \qquad 1 \le i < n\,.}
\end{equation}
Indeed, one can easily check that the set of (non-trivial) cluster variables in~\eqref{eq: A_n cluster variables} precisely corresponds to the symbol letters \eqref{eq:chainsym}, with a set of spurious letters of the form $-2Y_i$ ($i=1,\ldots,n-1$), which do not appear in the final expression for the symbol.

\subsection{Cluster adjacency}
\label{sec: cluster adj chain}

A symbol whose alphabet is the set of cluster variables of some cluster algebra is said to
satisfy \emph{cluster adjacency} if two variables appear next to each other in any term of the symbol
only if they are compatible in the mathematical sense reviewed in Section~\ref{sec:rev_cluster}.
This phenomenon, originally described in~\cite{Drummond:2017ssj}, has by now appeared in several
contexts where multiple polylogarithms arise in physics.  Here we show that the symbols
of de Sitter wavefunction coefficients associated with the $n$-chain graph satisfy cluster
adjacency with respect to the $A_{2n{-}2}$ cluster algebra.
The key aspect of the proof comes from the fact that the kinematic flow restricts
adjacent entries in the symbol to be associated with pairs of tubings
that are either nested (as in the left panel of
Figure~\ref{fig: examples tubings chain}) or disjoint. We now show that
under the map~(\ref{eq:cluster map chain}), this is equivalent to the non-crossing of
chords in the $(n+3)$-gon, which determines cluster compatibility of the corresponding cluster variables.

Indeed, from \eqref{eq:cluster map chain} one can check that boundary chords are associated with tubings enclosing either a single external or internal energy variable:
\begin{equation}
    \frac{X_{i,i}^{++}}{X} = \Delta_{2i-1,\,2i}\,,\quad 1 \le i \le n\,,\qquad \frac{-2Y_i}{X} = \Delta_{2i,2i+1}\,,\quad 1 \le i < n\,.
\end{equation}
Thus, region variables $X_{i,j}^{++}$ are mapped to chords as:
\begin{equation}
    \frac{X_{i,j}^{++}}{X} = \frac{1}{X}\left(\sum_{k=i}^{j}X_{k,k}^{++} - \sum_{k=i}^{j-1}2Y_k \right)= \sum_{k=2i-1}^{2j-1}\Delta_{k,k+1} = \Delta_{2i-1,\,2j}\,,\quad 1\leq i\leq j\leq n\,.
\end{equation}
In the same vein, the rest are given by:
\begin{equation}
\begin{aligned}
    \frac{X_{i,j}^{+-}}{X} &= \Delta_{2i-1,\,2j+1}\,,\quad &\frac{X_{i,j}^{-+}}{X} &= \Delta_{2i-2,\,2j}\,,\quad &\frac{X_{i,j}^{--}}{X} &= \Delta_{2i-2,2j+1}\,,\\
    \frac{X_{i,n}^{+\pm}}{X} &= \Delta_{2i-1,\,2n}\,,\quad &\frac{X_{1,j}^{\pm+}}{X} &= \Delta_{1,\,2j}\,,\quad &\frac{X_{1,n}^{\pm\pm}}{X} &= 1\,,
\end{aligned}
\end{equation}
where $1<i\leq j<n$. In other words, given a tube in the graph, the corresponding cluster variable is given by summing over the cluster variables associated with individual boundary chords of each enclosed node or marking. Graphically, this can be represented as e.g.:
\begin{equation}
\begin{aligned}
\resizebox{0.85\textwidth}{!}{
    \begin{tikzpicture}[baseline={(0,-0.1)}]
        
        
        \draw[thick] (-0.8, 0) -- (2, 0);
        \draw[thick] (2.7, 0) -- (4.8, 0);

        \draw[thick] (0, 0.25) -- (4.0, 0.25);
        \draw[thick] (0, -0.25) -- (4.0, -0.25);
        \draw[thick] (0, 0.25) arc (90:270:0.25);
        \draw[thick] (4.0, -0.25) arc (-90:90:0.25);

        \filldraw (0,0) circle (2pt);
        \node at (0, -0.6) {$i$};
        
        \node at (0.8,0) {$\times$};
        
        \filldraw (1.6,0) circle (2pt);
        \node at (1.6, -0.6) {$i+1$};
        
        \node at (2.4,0) {$\dots$};
        
        \filldraw (3.2,0) circle (2pt);
        \node at (3.2, -0.6) {$j$};
        
        \node at (4.0,0) {$\times$};

        \node[anchor=base] at (2.0, 0.8) {$X_{i,j}^{+-}/X$};

        \node at (5.5, 0) {$=$};

        
        \draw[thick] (6.2, 0) -- (9, 0);
        \draw[thick] (9.8, 0) -- (11.8, 0);

        \filldraw (7.0,0) circle (2pt);
        \node at (7.0, -0.6) {$i$};
        \draw[thick] (7.0,0) circle (0.2);

        \node at (7.8,0) {$\times$};
        \draw[thick] (7.8,0) circle (0.2);

        \filldraw (8.6,0) circle (2pt);
        \node at (8.6, -0.6) {$i+1$};
        \draw[thick] (8.6,0) circle (0.2);

        \node at (9.4,0) {$\dots$};
        
        \filldraw (10.2,0) circle (2pt);
        \node at (10.2, -0.6) {$j$};
        \draw[thick] (10.2,0) circle (0.2);

        \node at (11,0) {$\times$};
        \draw[thick] (11,0) circle (0.2);

    \end{tikzpicture} 
    }\\
    = \Delta_{2i-1,2i} + \Delta_{2i,2i+1}+\Delta_{2i+1,2i+2}+\ldots+\Delta_{2j-1,2j}+\Delta_{2j,2j+1}=\Delta_{2i-1,2j+1}\,.
\end{aligned}\label{eq:bdychords}
\end{equation}
Therefore, an overlap of two tubes will correspond to two intersecting chords in the $(n{+}3)$-gon:
\begin{equation}
    \resizebox{0.85\textwidth}{!}{
    \begin{tikzpicture}[baseline={(0,-0.1)}]
        
        
        \draw[thick] (0, 0) -- (7.8, 0);

        \filldraw (0,0) circle (2.5pt);
        \node at (0, -0.6) {$1$};
        
        \node at (1,0) {\large{$\times$}};
        
        \filldraw (2,0) circle (2.5pt);
        \node at (2, -0.6) {$2$};
        
        \node at (3,0) {$\times$};
        
        \filldraw (4,0) circle (2.5pt);
        \node at (4, -0.6) {$3$};
        
        \node at (5,0) {$\times$};
        
        \filldraw (6,0) circle (2.5pt);
        \node at (6, -0.6) {$4$};
        
        \node at (7,0) {$\times$};

        \node at (8.1,0) {$\dots$};

        \draw[red,thick] (-0.2, 0.3) -- (6.4, 0.3);
        \draw[red,thick] (-0.2, -0.3) -- (6.4, -0.3);
        \draw[red,thick] (-0.2, 0.3) arc (90:270:0.3);
        \draw[red,thick] (6.4, -0.3) arc (-90:90:0.3);

        \draw[blue,thick] (1.7, 0.22) -- (7.8, 0.22);
        \draw[blue,thick] (1.7, -0.22) -- (7.8, -0.22);
        \draw[blue,thick] (1.7, 0.22) arc (90:270:0.22);

        \node at (9, 0) {\Large $\longleftrightarrow$};

        
        \coordinate (C) at (13,0);

        \draw[thick] ($(C)+(60:2.5)$) arc (60:240:2.5);
        \draw[thick, loosely dotted] ($(C)+(240:2.5)$) arc (240:255:2.5);

        \coordinate (P1) at ($(C)+(60:2.5)$);
        \coordinate (D1) at ($(C)+(45:2)$);
        \coordinate (P2) at ($(C)+(90:2.5)$);
        \coordinate (P3) at ($(C)+(120:2.5)$);
        \coordinate (D3) at ($(C)+(135:1.8)$);
        \coordinate (P4) at ($(C)+(150:2.5)$);
        \coordinate (P5) at ($(C)+(180:2.5)$);
        \coordinate (P6) at ($(C)+(210:2.5)$);
        \coordinate (P7) at ($(C)+(240:2.5)$);

        \draw[red,thick] (P1) -- (P7);

        \coordinate (T) at ($(C)+(-60:2.2)$);
        \draw[blue,thick] (P3) -- (T);
        \draw[thick, loosely dotted] (T) -- ($(C)+(-60:3)$);

        \filldraw (P1) circle (2.5pt) node[above right] {$1$};
        \node at (D1) {$\Delta_{1,7}$};
        \filldraw (P2) circle (2.5pt) node[above left] {$2$};
        \filldraw (P3) circle (2.5pt) node[above left] {$3$};
        \node at (D3) {$\Delta_{3,k}$};
        \filldraw (P4) circle (2.5pt) node[left] {$4$};
        \filldraw (P5) circle (2.5pt) node[left] {$5$};
        \filldraw (P6) circle (2.5pt) node[left] {$6$};
        \filldraw (P7) circle (2.5pt) node[below left] {$7$};

    \end{tikzpicture}
    },
\end{equation}
and vice versa, which implies that cluster adjacency follows immediately from the rules of kinematic flow, and thus is automatically satisfied by the symbol.

\subsection{Example: 2-site chain}
\label{sec:chain2}

We start with the simplest case of a 2-site chain graph, where the symbol is explicitly given by \cite{Hillman:2019wgh}:
\begin{equation}\label{eq: symbol 2chain}
    \mathcal{S}_2^{(\text{chain})}=\frac{X_1+Y_1}{X_1+X_2}\otimes\frac{X_2-Y_1}{X_2+Y_1}+\frac{X_2+Y_1}{X_1+X_2}\otimes\frac{X_1-Y_1}{X_1+Y_1}\,.
\end{equation}
Here we recognize the entries to be formed by ratios of region variables, each associated with a certain tubing in the graph:
\begin{multline}\label{eq: 2chain tubings}
    \tikzset{
        base/.style={baseline=-0.6ex, scale=1},
        dot/.style={circle, fill=black, inner sep=1.5pt, outer sep=0pt},
        cross/.style={font=\bfseries\scriptsize, inner sep=0pt},
        tube/.style={draw, thick, rounded corners=3pt, line width=0.8pt} 
    }  
    \newcommand{\basegraph}{
        \draw[thick] (0,0) -- (0.8,0);
        \node[dot] (L) at (0,0) {};
        \node[dot] (R) at (0.8,0) {};
        \node[cross] (C) at (0.4,0) {$\times$};
    }
    \left\{ 
    \begin{tikzpicture}[base]
        \basegraph
        \draw[tube] (-0.15, -0.12) rectangle (0.15, 0.12);
    \end{tikzpicture}
    , 
    \begin{tikzpicture}[base]
        \basegraph
        \draw[tube] (-0.15, -0.12) rectangle (0.55, 0.12);
    \end{tikzpicture}
    , 
    \begin{tikzpicture}[base]
        \basegraph
        \draw[tube] (0.65, -0.12) rectangle (0.95, 0.12);
    \end{tikzpicture}
    , 
    \begin{tikzpicture}[base]
        \basegraph
        \draw[tube] (0.25, -0.12) rectangle (0.95, 0.12);
    \end{tikzpicture}
    , 
    \begin{tikzpicture}[base]
        \basegraph
        \draw[tube] (-0.15, -0.15) rectangle (0.95, 0.15);
    \end{tikzpicture}
    \right\}=
    \left\{ X_1{+}Y_1, \, X_1{-}Y_1, \, X_2{+}Y_1, \, X_2{-}Y_1, \, X_1{+}X_2 \right\}.
\end{multline}
This symbol alphabet can be reproduced with the following map to the cluster~(\ref{eq: A type matrix param})
of the $A_{2n{-}2} = A_2$ cluster algebra:
\begin{equation}
\label{eq:2param}
    z_1 = \frac{X_1+Y_1}{X_1+X_2}\,,\quad z_2 = \frac{X_1-Y_1}{X_1+X_2}\,.
\end{equation}
Note that in this parameterization, the full symbol alphabet of the cluster algebra
\begin{align}\label{eq: A_2 variables}
    \{ z_1,\, z_2,\, 1{-}z_1,\,1{-}z_2,\,z_2-z_1\}
    =\left\{\frac{X_1+Y_1}{X_1+X_2},\,\frac{X_1-Y_1}{X_1+X_2},\, \frac{X_2-Y_1}{X_1+X_2},\,\frac{X_2+Y_1}{X_1+X_2}, \frac{-2Y_1}{X_1+X_2} \right\}
\end{align}
contains all of the letters appearing in the symbol together with a fifth spurious letter.
Finally,
in terms of the Pl\"ucker variables associated with the chords $\Delta_{i,j}$ (see Figure~\ref{fig: A_2 pentagon}), the symbol~(\ref{eq: symbol 2chain})
then takes the form
\begin{align}
\label{eq:s2symb}
\mathcal{S}_2^\text{(chain)} = \Delta_{1,2} \otimes \Delta_{2,4} + \Delta_{3,4} \otimes \Delta_{1,3} - \Delta_{1,2} \otimes \Delta_{3,4} - \Delta_{3,4} \otimes \Delta_{1,2}
\end{align}
which manifests the property of cluster adjacency as adjacent terms in the symbol involve non-crossing chords.

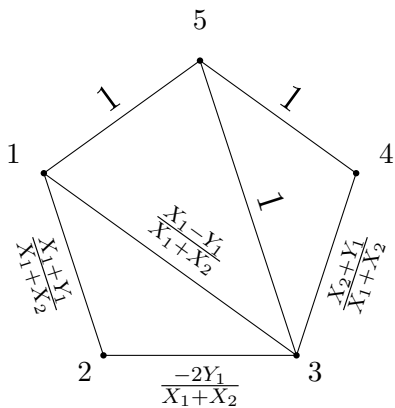
\begin{figure}
\centering
    \begin{tikzcd}
    \begin{tikzpicture}[scale=1.2, baseline=(current bounding box.center)]

    \def\R{1.8}
    \foreach \i/\lab/\ang in {
            1/5/90, 2/1/162, 3/2/234, 4/3/306, 
            5/4/18
    } 
    {
            \coordinate (v\i) at (\ang:\R);
            \fill (v\i) circle (1pt);
            \node at (\ang:\R+0.35) {$\lab$};
    }
    
    \draw
        (v1) -- node[midway, sloped, above] {\Large{1}} (v2)
        -- node[midway, sloped, below] {$\frac{X_1+Y_1}{X_1+X_2}$} (v3)
        -- node[midway, sloped, below] {$\frac{-2Y_1}{X_1+X_2}$} (v4)
        -- node[midway, sloped, below] {$\frac{X_2+Y_1}{X_1+X_2}$} (v5)
        -- node[midway, sloped, above] {\Large{1}} (v1);
        \draw
        (v2) -- node[midway, sloped, above] {$\frac{X_1-Y_1}{X_1+X_2}$} (v4);
        \draw
        (v1) -- node[midway, sloped, above] {\Large{1}} (v4);
        
    \end{tikzpicture}
    \end{tikzcd}
    \caption{Example of a triangulation of the pentagon associated with the $A_2$ cluster algebra of the 2-site chain. Each chord $\Delta_{i,j}$ has been labeled by its associated Pl\"ucker variable obtained from plugging~(\ref{eq:2param}) into~(\ref{eq: A type matrix param}).}
    \label{fig: A_2 pentagon}
    \end{figure}

\subsection{Example: 3-site chain}
\label{sec:chain3}

The next simplest example is the 3-site chain graph, for which the symbol alphabet consists of the 13 letters
\begin{multline}\label{eq: alphabet 3chain}
    \{X_1-Y_1,\, X_1+Y_1,\, X_3-Y_2,\, X_3+Y_2,\, X_2+X_3-Y_1,\, X_2+X_3+Y_1,\, X_1+X_2-Y_2,\\
    X_1+X_2+Y_2,\, X_2-Y_1-Y_2,\, X_2+Y_1-Y_2,\, X_2-Y_1+Y_2,\, X_2+Y_1+Y_2,\, X_1+X_2+X_3\},
\end{multline}
each of which corresponds to some tubing of the graph.
In this case the map~\eqref{eq:cluster map chain} to the $A_{2n{-}2} = A_4$ cluster algebra is given by~(\ref{eq: A type matrix param}) with
\begin{equation}
    z_1 = \frac{X_1+Y_1}{X_1+X_2+X_3}\,,\ z_2 = \frac{X_1-Y_1}{X_1+X_2+X_3}\,,\ z_3 = \frac{X_1+X_2+Y_2}{X_1+X_2+X_3}\,,\ z_4 = \frac{X_1+X_2-Y_2}{X_1+X_2+X_3}\,,
\end{equation}
and all cluster variables take the form $z_i$, $1-z_i$ or $z_{i,j}=z_i-z_j$. One can check that the resulting set contains all letters in \eqref{eq: alphabet 3chain}, as well as two spurious letters $\{-2Y_1,\, -2Y_2\}$ that do not appear in the symbol.

To check cluster adjacency we identify the different variables with chords $\Delta_{i,j}$ in a heptagon. As before, two chords that cross each other will never be part of the same triangulation, and thus cannot appear as adjacent entries in the symbol. Since crossing chords are associated with pairs of tubes that are neither nested nor disjoint, this is consistent with the rules of kinematic flow, showing that the symbol indeed satisfies cluster adjacency.
For instance, consider the chord $\Delta_{1,3}$, which is mapped to:
\begin{equation}
\label{eq:example}
    \Delta_{1,3} = z_2 = \frac{X_1-Y_1}{X_1+X_2+X_3} = \frac{X_{11}^{+-}}{X_1+X_2+X_3} = \tikzset{
        base/.style={baseline=-0.6ex, scale=1},
        dot/.style={circle, fill=black, inner sep=1.5pt, outer sep=0pt},
        cross/.style={font=\bfseries\scriptsize, inner sep=0pt},
        tube/.style={draw, thick, rounded corners=4.5pt, line width=0.8pt},
        chain/.style={thick}
    }
    \newcommand{\drawchain}{
        \draw[chain] (0,0) -- (3.2,0);
        \node[dot] at (0,0) {};    
        \node[cross] at (0.8,0) {$\times$}; 
        \node[dot] at (1.6,0) {};  
        \node[cross] at (2.4,0) {$\times$}; 
        \node[dot] at (3.2,0) {};  
    }
    \begin{tikzpicture}[base]
        \drawchain
        \draw[tube] (-0.2, -0.16) rectangle (1, 0.16);
    \end{tikzpicture}\,.
\end{equation}
The only three non-trivial chords that cross it are:
\begin{equation}
\begin{aligned}
    \Delta_{2,4} &= z_{3,1} = \frac{X_2-Y_1+Y_2}{X_1+X_2+X_3} = \frac{X_{22}^{-+}}{X_1+X_2+X_3}= \tikzset{
        base/.style={baseline=-0.6ex, scale=1},
        dot/.style={circle, fill=black, inner sep=1.5pt, outer sep=0pt},
        cross/.style={font=\bfseries\scriptsize, inner sep=0pt},
        tube/.style={draw, thick, rounded corners=4.5pt, line width=0.8pt},
        chain/.style={thick}
    }
    \newcommand{\drawchain}{
        \draw[chain] (0,0) -- (3.2,0);
        \node[dot] at (0,0) {};    
        \node[cross] at (0.8,0) {$\times$}; 
        \node[dot] at (1.6,0) {};  
        \node[cross] at (2.4,0) {$\times$}; 
        \node[dot] at (3.2,0) {};  
    }
    \begin{tikzpicture}[base]
        \drawchain
        \draw[tube] (0.6, -0.16) rectangle (1.9, 0.16);
    \end{tikzpicture}\,,\\ \Delta_{2,5} &= z_{4,1} = \frac{X_2-Y_1-Y_2}{X_1+X_2+X_3} = \frac{X_{22}^{--}}{X_1+X_2+X_3}=\tikzset{
        base/.style={baseline=-0.6ex, scale=1},
        dot/.style={circle, fill=black, inner sep=1.5pt, outer sep=0pt},
        cross/.style={font=\bfseries\scriptsize, inner sep=0pt},
        tube/.style={draw, thick, rounded corners=4.5pt, line width=0.8pt},
        chain/.style={thick}
    }
    \newcommand{\drawchain}{
        \draw[chain] (0,0) -- (3.2,0);
        \node[dot] at (0,0) {};    
        \node[cross] at (0.8,0) {$\times$}; 
        \node[dot] at (1.6,0) {};  
        \node[cross] at (2.4,0) {$\times$}; 
        \node[dot] at (3.2,0) {};  
    }
    \begin{tikzpicture}[base]
        \drawchain
        \draw[tube] (0.6, -0.16) rectangle (2.6, 0.16);
    \end{tikzpicture}\,,\\ \Delta_{2,6} &= 1-z_1 = \frac{X_2+X_3-Y_1}{X_1+X_2+X_3} = \frac{X_{2,3}^{-+}}{X_1+X_2+X_3}=\tikzset{
        base/.style={baseline=-0.6ex, scale=1},
        dot/.style={circle, fill=black, inner sep=1.5pt, outer sep=0pt},
        cross/.style={font=\bfseries\scriptsize, inner sep=0pt},
        tube/.style={draw, thick, rounded corners=4.5pt, line width=0.8pt},
        chain/.style={thick}
    }
    \newcommand{\drawchain}{
        \draw[chain] (0,0) -- (3.2,0);
        \node[dot] at (0,0) {};    
        \node[cross] at (0.8,0) {$\times$}; 
        \node[dot] at (1.6,0) {};  
        \node[cross] at (2.4,0) {$\times$}; 
        \node[dot] at (3.2,0) {};  
    }
    \begin{tikzpicture}[base]
        \drawchain
        \draw[tube] (0.6, -0.16) rectangle (3.4, 0.16);
    \end{tikzpicture}\,,
\end{aligned}
\end{equation}
which are precisely the tubes that overlap with the one shown in~(\ref{eq:example}), and the corresponding entries
do not appear in adjacent entries of the symbol.

\subsection{Bootstrap}
\label{sec:chainboot}

Although it is of course well-known how to compute the symbol of de Sitter wavefunction coefficients
via various methods,
it is interesting to identify a set of properties that would allow them to be bootstrapped, following
successful examples of this paradigm in other contexts reviewed in the introduction.
We therefore consider a bootstrap program based on the following inputs:
\begin{enumerate}
    \item Symbol alphabet: drawn from the set of $A_{2n{-}2}$ cluster variables under the identification
    defined by (\ref{eq: A type matrix param}) and (\ref{eq:cluster map chain}).
    \item Integrability: a linear combination $\sum_{i_1,\dots,i_n} c_{i_1,\dots,i_n}[\Phi_{i_1}\otimes\cdots\otimes\Phi_{i_n}]$ of words is the symbol of a function if and only if
    \begin{equation}
    \begin{aligned}
        \sum_{i_1,\cdots,i_n} c_{i_1,\cdots,i_n}\,[\Phi_{i_1}{\otimes}\cdots{\otimes}\Phi_{i_{j{-}1}}{\otimes}\Phi_{i_{j{+}2}}\cdots{\otimes}\Phi_{i_n}]\ \text{dlog}\Phi_{i_j}\wedge\text{dlog}\Phi_{i_{j{+}1}}=0
    \end{aligned}
    \end{equation}
    for all $j \in \{1,\ldots,n-1\}$. This is simply the statement that partial derivatives acting on a function must commute.

    \item First entry condition: the first entries of each term in the symbol determine its branch points on the physical sheet, and must therefore correspond to physical singularities~\cite{Gaiotto:2011dt}. In terms of region variables, this means that the first entries can only take the form $X_{i,j}^{++}$. 
    
    \item Cluster adjacency: neighboring symbol entries must be cluster compatible, i.e.~the chords corresponding to neighboring symbol entries do not cross each other.

    \item Discrete symmetry: the symbol of the $n$-site chain is invariant under flip symmetry
    \begin{equation}
    \label{eq:flip}
        X_{i} \leftrightarrow X_{n-i+1} \quad \text{and}\quad Y_{i} \leftrightarrow Y_{n-i}.
    \end{equation}

    \item Soft limit: the symbol of the wave function coefficient vanishes if we send any internal energy to zero $Y_i\to 0$. This condition arises because the bulk-to-bulk propagator is dressed with an internal energy factor in the numerator so that the integrand becomes a canonical form; see the discussion in Appendix D of~\cite{Arkani-Hamed:2023kig} for more details. In particular, it is important to note that all of our discussion applies to $\psi \times \left(\prod_i Y_i\right)$, where $\psi$ is the wavefunction coefficient and this manifestly has a soft limit when any $Y_i\to 0$.

    \item Spurious letters: in principle one should also impose that the spurious letters drop out, but in all of the examples we have checked this happens automatically after all of the above constraints are imposed.
    
\end{enumerate}

Let us demonstrate the 2-site chain bootstrap in detail. The first step is to impose the integrability condition on the $5\times5$ combination of weight-two symbols made of the following letters: 
\begin{align}
    \Phi_{1,\dots,5}&=\left\{\frac{X_1+Y_1}{X_1+X_2},\,\frac{X_1-Y_1}{X_1+X_2},\, \frac{X_2-Y_1}{X_1+X_2},\,\frac{X_2+Y_1}{X_1+X_2}, \frac{-2Y_1}{X_1+X_2} \right\}
\\
&= \{ \Delta_{1,2}, \Delta_{1,3}, \Delta_{2,4}, \Delta_{3,4}, \Delta_{2,3}\}\,.
\label{eq:todo}
\end{align}
We find that there are 19 integrable weight-two symbols. Next, the first entry condition only allows $\Delta_{1,2}$ and $\Delta_{3,4}$
to appear as the first entry. This constraint reduces the symbol space to five dimensions, a basis for which is given by
\begin{equation}
    \{ \Delta_{3,4} \otimes \Delta_{3,4},~~ \Delta_{3,4} \otimes \Delta_{1,3}, ~~\Delta_{1,2} \otimes \Delta_{3,4}
    + \Delta_{3,4} \otimes \Delta_{1,2},~~ \Delta_{1,2} \otimes \Delta_{2,4},~~ \Delta_{1,2} \otimes \Delta_{1,2}\}\,.
\end{equation}
Notice that the spurious letter $\Delta_{2,3}$ has already dropped out. Also, it is interesting to note that each term already manifests cluster adjacency; imposing it does not further
constrain the space of our ansatz. However, the requirement that the symbol be invariant under the flip symmetry $X_1\leftrightarrow X_2$ cuts this down to the 3-dimensional space spanned by
\begin{equation}
\{ \Delta_{1,2} \otimes \Delta_{1,2} + \Delta_{3,4} \otimes \Delta_{3,4}, ~~
\Delta_{1,2} \otimes \Delta_{2,4} + \Delta_{3,4} \otimes \Delta_{1,3}, ~~
\Delta_{1,2} \otimes \Delta_{3,4} + \Delta_{3,4} \otimes \Delta_{1,2}\}\,.
\end{equation}
Finally, the soft limit fixes the symbol to be~(\ref{eq:s2symb}), uniquely up to an overall normalization.

It is interesting that the soft limit 6 (along with bootstrap constraints 1, 2, 3 and 5) is enough to fix the symbol of the 2-site chain; in this sense, we can say that the other conditions collectively imply that the symbol must satisfy cluster adjacency (and that the spurious letter drops out).
We have checked that this property persists for the 3- and 4-site chain graphs.
It would be interesting to see if it remains true for all $n$.
We summarize the dimensions of the symbol spaces at each step in the cluster bootstrap for the 2-, 3- and 4-site chains in Table~\ref{tab: chain summary}. For details on implementing the symbol bootstrap algorithm, we refer the reader to Appendix B of~\cite{Dixon:2016nkn}. The null-space computation package~\cite{Li_SparseRREF} was utilized for these calculations.

\begin{table}
\begin{longtable}{|C{5cm}|C{1.cm}|C{1.cm}|C{1.cm}|C{1.cm}|C{1.cm}|C{1.cm}|}
    \hline
    Constraints
    & \multicolumn{2}{c|}{2-site chain}
    & \multicolumn{2}{c|}{3-site chain}
    & \multicolumn{2}{c|}{4-site chain} \\
    \hline
    Integrability and first entry & 5 & 5 & 102 & 102 & 2736 & 2736 \\
    \hline
    Adjacency & 5 & N/I & 90 & N/I & 1810 & N/I\\
    \hline
    Flip Symmetry & 3 & 3 & 47 & 54 & 914 & 1385\\
    \hline
    $Y_i \to 0$ limit & 1 & 1 & 1 & 1 & 1 & 1 \\
    \hline
    \caption{A summary of the 2-, 3- and 4-site chain symbol bootstrap. The symbols have transcendental weight 2, 3 and 4, respectively and the initial alphabets have 5, 14, and 27 dimensionless letters. There are two columns for each case; the answer is uniquely determined (up to an overall constant) either with (left columns) or without (right columns) imposing cluster adjacency. The spurious letters drop out automatically without being imposed; this happens after imposing integrability and the first-entry condition for the 2-site chain, and after imposing the $Y_i\to 0$ limit constraint for 3- and 4-site chains.}
    \label{tab: chain summary}
\end{longtable}
\end{table}

Here we also provide a comparison of counting of ordered pairs of letters allowed by cluster compatibility and the number of ordered adjacent pairs that appear in the symbol of chain graphs. Consider the 2-site chain as a concrete example. There are 5 letters, which form 25 ordered pairs of letters. Cluster compatibility rules out two ordered pairs $\{(\Delta_{1,3},\Delta_{2,4}),(\Delta_{2,4},\Delta_{1,3})\}$ and we are left with 23 ordered pairs of letters. On the other hand, the final symbol of the 2-site chain wavefunction coefficient only contains 4 ordered pairs, namely $\{(\Delta_{1,2},\Delta_{2,4}),\ (\Delta_{1,2},\Delta_{3,4}),\ (\Delta_{3,4},\Delta_{1,3}),(\Delta_{3,4},\Delta_{1,2})\}$. The counting of higher-site chain graphs is summarized in Table \ref{tab: chain compatible counting}.
\begin{table}[h!]
    \centering
    \begin{tabular}{|c|c|c|}
        \hline
        Example & Cluster compatible ordered pairs & Wavefunction ordered pairs \\
        \hline
        2-site chain & 23 & 4\\
        \hline
        3-site chain & 166 & 52\\
        \hline
        4-site chain & 589 & 216\\
        \hline
    \end{tabular}
    \caption{A counting of cluster compatible letter pairs and pairs that appear in the chain wavefunction coefficients.}
    \label{tab: chain compatible counting}
\end{table}

\section{Polygonal Graphs}
\label{sec:gon graphs}

In this section we explore
 the cluster algebra structure corresponding to $n$-gon graphs. In Section~\ref{sec:gonmap} we 
 demonstrate a map between the symbol 
alphabet of the wavefunction coefficient associated with the $n$-site loop, which
is a weight-$n$ multiple polylogarithm function, and the cluster variables of the $B_{2n{-}1}$ cluster algebra.
In Section \ref{sec:gonadj}
we demonstrate that the symbols satisfy
cluster adjacency with respect to this algebra.
Then, after a few explicit examples in Sections \ref{sec:gon2} and \ref{sec:gon3}, we discuss
the cluster bootstrap for computing these symbols in Section~\ref{sec:gonboot}.

\subsection{Symbol alphabets from \texorpdfstring{$B$}{B}-type cluster algebras}
\label{sec:gonmap}

Consider a loop graph with $n$ sites. The tubes of this graph determine its symbol alphabet:
\begin{align}\label{eq:gonsym}
    X_{j,k}^{\pm\pm}&=\sum_{i=j}^k X_i \pm Y_{j{-}1}\pm Y_k\,,\quad |k-j|< n-1\,,\qquad X = \sum_{i=1}^n X_i\,,\nonumber\\
    X_{j+1,j}^{++}&=\sum_{i=1}^{n} X_i + 2 Y_{j}\,,\quad 1\le j\le n\,,
\end{align}
with the understanding that the labels are identified mod $n$. This gives a total of $4n(n-1)+n+1$ variables and a symbol alphabet of $n(4n-3)$ dimensionless ratios.

For this $n$-gon graph, we define our map from the symbol alphabet \eqref{eq:gonsym} to the cluster variables of $B_{2n{-}1}$ as follows:
\begin{equation}\label{eq: map n-gon}
    \boxed{z_{2a,\,2a-1} = \frac{1}{X}\left(X_a+Y_{a-1}+Y_a \right),\ \  1\leq a \leq n,\ \quad z_{2a+1,\,2a} = \frac{-2Y_a}{X},\ \ 1\leq a < n\,,}
\end{equation}
where we have specified differences of consecutive $z_a$ instead of the $z_a$ themselves, since all cluster variables can be written in terms of the former. This gives a total of $2n(2n{-}1)$ letters.

Let us now check that the assignment~\eqref{eq: map n-gon} indeed reproduces the complete set of region variables corresponding to the tubes we can draw in an $n$-gon graph. If we restrict ourselves to cluster variables $\Delta_{a,b}$ with no barred indices for the moment, we see that: 
\begin{equation}
    \begin{aligned}
        &\frac{X_{a,b}^{+-}}{X} = \Delta_{2a-1,\,2b+1}\,,\quad &\frac{X_{a,b}^{++}}{X} &= \Delta_{2a-1,\,2b}\,,\\
        &\frac{X_{a,b}^{--}}{X} = \Delta_{2a-2,2b+1}\,,\quad &\frac{X_{a,b}^{-+}}{X} &= \Delta_{2a-2,2b}\,.
    \end{aligned}
\end{equation}
However, since the indices $a,b$ are strictly ordered such that $a<b$, we are missing half of the region variables that can appear in the symbol alphabet. The missing ones are precisely $\Delta_{a,\bar b}$, since the corresponding tubings are obtained as the complements of the previous ones. In other words, we have successfully mapped the set of symbol letters for an $n$-gon graph to the cluster variables of a $B_{2n{-}1}$ algebra.
Similarly to the case of the $n$-chain graph, the latter had $n$ additional spurious letters $-2 Y_i$ that do not appear in the symbol of the wavefunction coefficient.

\subsection{Cluster adjacency}
\label{sec:gonadj}
Following a similar logic to our argument for chain graphs, we show that the notion of cluster compatibility follows immediately from the rules of kinematic flow, and thus the symbol for any $n$-gon graph satisfies cluster adjacency. The reasoning is completely analogous to the one presented in Section \ref{sec: cluster adj chain}: the only difference is that now each cluster variable is associated with two chords (since one has to include the mirror image), and thus the four chords associated with a pair of cluster variables have to be mutually non-crossing for the corresponding tubes to be compatible. 
We present some concrete examples for the bubble and triangle graphs in the following sections.

\subsection{Example: 2-site loop}
\label{sec:gon2}
We start with the 2-site loop, or bubble graph, for which the symbol can be explicitly written as \cite{Hillman:2019wgh}:
\begin{equation}\label{eq: symbol bubble}
\begin{aligned}
    \mathcal{S}_2^{(\text{loop})} = &-(1-u_1)\otimes\frac{u_1}{1-v_1}-(1-v_1)\otimes\frac{v_1}{1-u_1} \\ &-(1-u_2)\otimes\frac{u_2}{1-v_2}-(1-v_2)\otimes\frac{v_2}{1-u_2}\\ &+(1-u_3)\otimes\frac{u_3}{1-v_3}+(1-v_3)\otimes\frac{v_3}{1-u_3}
\end{aligned}
\end{equation}
in terms of the cross-ratios
\begin{equation}
\begin{aligned}
    u_1&=\frac{X_1+Y_1-Y_2}{X_1+X_2+2Y_1}\,,\quad &u_2&=\frac{X_1+Y_2-Y_1}{X_1+X_2+2Y_2}\,,\quad &u_3&=\frac{X_1-Y_1-Y_2}{X_1+X_2}\,,\\ v_1&=\frac{X_2+Y_1-Y_2}{X_1+X_2+2Y_1}\,,\quad &v_2&=\frac{X_2+Y_2-Y_1}{X_1+X_2+2Y_2}\,,\quad &v_3&=\frac{X_2-Y_1-Y_2}{X_1+X_2}\,.
\end{aligned}
\end{equation}
Once we expand all of the $1{-}u_i$ and $1{-}v_i$, we can see that the symbol alphabet is composed of only 11 independent (dimensionful) letters:
\begin{multline}
    \{X_1+X_2,\,X_1+Y_1+Y_2,\,X_2+Y_1+Y_2,\,X_1-Y_1+Y_2,\,X_2-Y_1+Y_2,\,X_1+Y_1-Y_2,\\X_2+Y_1-Y_2,\,X_1-Y_1-Y_2,\,X_2-Y_1-Y_2,\,X_1+X_2+2Y_1,\,X_1+X_2+2Y_2\}.
\end{multline}
This symbol alphabet can be expressed in terms of cluster variables using the map~\eqref{eq: map n-gon}:
\begin{equation}
\label{eq:delta1}
    \Delta_{1,2}=z_{2,1} = \frac{X_1+Y_1+Y_2}{X_1+X_2}\,,\quad \Delta_{2,3} = z_{3,2} = \frac{-2Y_1}{X_1+X_2}\,,\quad \Delta_{3,4} = z_{4,3} = \frac{X_2+Y_1+Y_2}{X_1+X_2}\,.
\end{equation}
The remainder of the unbarred variables can be constructed simply via
\begin{equation}\label{eq: cluster vars from boundary chords}
    \Delta_{a,b} = \Delta_{a,a+1} + \Delta_{a+1,a+2} + \ldots + \Delta_{b-1,b}\,,
\end{equation}
and the ones involving barred indices will be given by $\Delta_{a,\bar b} = 1-\Delta_{a,b}$. One can easily check that the set of all cluster variables indeed reproduces the symbol alphabet, along with two additional spurious letters $-2Y_1$ and $-2Y_2$.

In order to check cluster adjacency, we place the different cluster variables as chords in an octagon with vertices labeled $\{1,2,3,4,\bar 1,\bar2,\bar3,\bar4\}$ as in Figure \ref{fig: crossing bar 2,3}. Two cluster variables are compatible if they appear together in a symmetric triangulation of this octagon (as defined in Section \ref{sec:rev_cluster}). Equivalently, two variables are compatible only if the corresponding set of four chords (for each variable, we need to include the mirror image under $a\leftrightarrow\bar a$) are non-crossing.

Similar to the logic for the chain graphs, this condition turns out to be completely equivalent to the tubes of the two associated region variables being either nested or disjoint. As an example, consider the cluster variable:
\begin{equation}
    \Delta_{2,\bar 3}= \frac{X_1+X_2+2Y_1}{X_1+X_2}=\vcenter{\hbox{\begin{tikzpicture}[
    scale=0.3, , 
    every node/.style={font=\small}
]
    \coordinate (X1) at (-2, 0);
    \coordinate (X2) at (2, 0);
    \coordinate (Y1) at (0, 1);
    \coordinate (Y2) at (0, -1);
    \draw[line width=1.2pt] (0,0) ellipse (2 and 1);
    \fill (X1) circle (5pt);
    \fill (X2) circle (5pt);
    \node[font=\tiny] at (Y1) {$\times$};
    \node[font=\tiny] at (Y2) {$\times$};
    \draw[line width=1pt] (-1.6, 0.2)
        .. controls (-1.6, 0.6) and (-2.4, 0.5) .. (-2.4, 0)      
        .. controls (-2.4, -0.8) and (-1.3, -1.35) .. (0, -1.35)  
        .. controls (1.3, -1.35) and (2.4, -0.8) .. (2.4, 0)      
        .. controls (2.4, 0.5) and (1.6, 0.6) .. (1.6, 0.2)       
        .. controls (1.5, -0.3) and (0.9, -0.65) .. (0, -0.65)    
        .. controls (-0.9, -0.65) and (-1.5, -0.3) .. (-1.6, 0.2); 

    \node[left] at (-2.5, 0) {$X_1$};
    \node[right] at (2.5, 0) {$X_2$};
    \node[above] at (0, 1.2) {$Y_1$};
    \node[below] at (0, -1.5) {$Y_2$};
\end{tikzpicture}}}.
\end{equation}
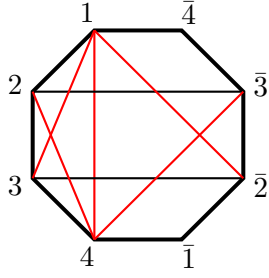
\begin{figure}
\centering    
\begin{tikzpicture}[
    scale=0.5,
    every node/.style={font=\normalsize},
    thick 
]

\def\R{3} 

\foreach \i/\lab/\ang in {
    1/1/112.5, 
    2/2/157.5, 
    3/3/202.5, 
    4/4/247.5, 
    5/\bar{1}/292.5, 
    6/\bar{2}/337.5, 
    7/\bar{3}/22.5, 
    8/\bar{4}/67.5
} {
    \coordinate (v\i) at (\ang:\R);
    \node at (\ang:\R+0.5) {$\lab$};
}

\draw[line width=1.5pt] (v1) -- (v2) -- (v3) -- (v4) -- (v5) -- (v6) -- (v7) -- (v8) -- cycle;

\draw[red] (v1) -- (v4); 
\draw (v2) -- (v7); 
\draw (v3) -- (v6); 

\draw[red] (v1) -- (v3); 
\draw[red] (v2) -- (v4); 

\draw[red] (v1) -- (v6); 
\draw[red] (v4) -- (v7); 

\end{tikzpicture}
\caption{Chords in the octagon that cross either $\Delta_{2,\bar 3}$ or $\Delta_{\bar 2,3}$. Here, we have drawn in red only one copy of each chord corresponding to a cluster variable that is not compatible with the ones shown in black.}
\label{fig: crossing bar 2,3}
\end{figure}
As can be seen in Figure \ref{fig: crossing bar 2,3}, the chords that cross either $\Delta_{2,\bar 3}$ or $\Delta_{\bar 2,3}$, and thus form the set of incompatible cluster variables, are:
\begin{align}
    \Delta_{1,3} &= \frac{X_1-Y_1+Y_2}{X_1+X_2} = \vcenter{\hbox{\begin{tikzpicture}[
    scale=0.3, 
    baseline={([yshift=15pt]current bounding box.center)}, 
    every node/.style={font=\small}
]
    \coordinate (X1) at (-2, 0);
    \coordinate (X2) at (2, 0);
    \coordinate (Y1) at (0, 1);
    \coordinate (Y2) at (0, -1);
    \draw[line width=1.2pt] (0,0) ellipse (2 and 1);
    \fill (X1) circle (5pt);
    \fill (X2) circle (5pt);
    \node[font=\tiny] at (Y1) {$\times$};
    \node[font=\tiny] at (Y2) {$\times$};
    \draw[line width=1pt] (-2.3, -0.1)
        .. controls (-2.5, 0.8) and (-1.0, 1.4) .. (0.1, 1.3)   
        .. controls (0.5, 1.3) and (0.5, 0.7) .. (0.1, 0.7)     
        .. controls (-0.8, 0.7) and (-1.5, 0.4) .. (-1.6, 0)    
        .. controls (-1.7, -0.4) and (-2.1, -0.4) .. (-2.3, -0.1); 
    \node[left] at (-2.5, 0) {$X_1$};
    \node[right] at (2.2, 0) {$X_2$};
    \node[above] at (0, 1.5) {$Y_1$};
    \node[below] at (0, -1.5) {$Y_2$};
\end{tikzpicture}}},\quad
\Delta_{1,4}  = \frac{X_1+X_2+2Y_2}{X_1+X_2}=\vcenter{\hbox{\begin{tikzpicture}[
    scale=0.3, 
    baseline={([yshift=15pt]current bounding box.center)}, 
    every node/.style={font=\small}
]
    \coordinate (X1) at (-2, 0);
    \coordinate (X2) at (2, 0);
    \coordinate (Y1) at (0, 1);
    \coordinate (Y2) at (0, -1);
    \draw[line width=1.2pt] (0,0) ellipse (2 and 1);
    \fill (X1) circle (5pt);
    \fill (X2) circle (5pt);
    \node[font=\tiny] at (Y1) {$\times$};
    \node[font=\tiny] at (Y2) {$\times$};
    \draw[line width=1pt] (-1.6, -0.2)
        .. controls (-1.6, -0.6) and (-2.4, -0.5) .. (-2.4, 0)      
        .. controls (-2.4, 0.8) and (-1.3, 1.35) .. (0, 1.35)       
        .. controls (1.3, 1.35) and (2.4, 0.8) .. (2.4, 0)          
        .. controls (2.4, -0.5) and (1.6, -0.6) .. (1.6, -0.2)      
        .. controls (1.5, 0.3) and (0.9, 0.65) .. (0, 0.65)         
        .. controls (-0.9, 0.65) and (-1.5, 0.3) .. (-1.6, -0.2);   
    \node[left] at (-2.5, 0) {$X_1$};
    \node[right] at (2.5, 0) {$X_2$};
    \node[above] at (0, 1.5) {$Y_1$};
    \node[below] at (0, -1.2) {$Y_2$};
\end{tikzpicture}}}\,\nonumber,\\
\Delta_{2,4}  &= \frac{X_2-Y_1+Y_2}{X_1+X_2}=\vcenter{\hbox{\begin{tikzpicture}[
    scale=0.3, 
    baseline={([yshift=15pt]current bounding box.center)}, 
    every node/.style={font=\small}
]
    \coordinate (X1) at (-2, 0);
    \coordinate (X2) at (2, 0);
    \coordinate (Y1) at (0, 1);
    \coordinate (Y2) at (0, -1);
    \draw[line width=1.2pt] (0,0) ellipse (2 and 1);
    \fill (X1) circle (5pt);
    \fill (X2) circle (5pt);
    \node[font=\tiny] at (Y1) {$\times$};
    \node[font=\tiny] at (Y2) {$\times$};
    \draw[line width=1pt] (2.3, -0.1)
        .. controls (2.5, 0.8) and (1.0, 1.4) .. (-0.1, 1.3)    
        .. controls (-0.5, 1.3) and (-0.5, 0.7) .. (-0.1, 0.7)  
        .. controls (0.8, 0.7) and (1.5, 0.4) .. (1.6, 0)       
        .. controls (1.7, -0.4) and (2.1, -0.4) .. (2.3, -0.1); 
    \node[left] at (-2.5, 0) {$X_1$};
    \node[right] at (2.5, 0) {$X_2$};
    \node[above] at (0, 1.5) {$Y_1$};
    \node[below] at (0, -1.2) {$Y_2$};
\end{tikzpicture}}},\quad
    \Delta_{1,\bar2} = \frac{X_2-Y_1-Y_2}{X_1+X_2}=\vcenter{\hbox{\begin{tikzpicture}[
    scale=0.3, 
    baseline={([yshift=15pt]current bounding box.center)}, 
    every node/.style={font=\small}
]
    \coordinate (X1) at (-2, 0);
    \coordinate (X2) at (2, 0);
    \coordinate (Y1) at (0, 1);
    \coordinate (Y2) at (0, -1);
    \draw[line width=1.2pt] (0,0) ellipse (2 and 1);
    \fill (X1) circle (5pt);
    \fill (X2) circle (5pt);
    \node[font=\tiny] at (Y1) {$\times$};
    \node[font=\tiny] at (Y2) {$\times$};
    \draw[line width=1pt] (2.4, 0)
        .. controls (2.4, 1.35) and (0.8, 1.35) .. (-0.1, 1.35)      
        .. controls (-0.6, 1.35) and (-0.6, 0.65) .. (-0.1, 0.65)    
        .. controls (0.8, 0.65) and (1.4, 0.65) .. (1.4, 0)          
        .. controls (1.4, -0.65) and (0.8, -0.65) .. (-0.1, -0.65)   
        .. controls (-0.6, -0.65) and (-0.6, -1.35) .. (-0.1, -1.35) 
        .. controls (0.8, -1.35) and (2.4, -1.35) .. (2.4, 0);       
    \node[left] at (-2.5, 0) {$X_1$};
    \node[right] at (2.5, 0) {$X_2$};
    \node[above] at (0, 1.5) {$Y_1$};
    \node[below] at (0, -1.5) {$Y_2$};
\end{tikzpicture}}}\,,\nonumber\\
\Delta_{3,\bar4} &= \frac{X_1-Y_1-Y_2}{X_1+X_2} =\vcenter{\hbox{\begin{tikzpicture}[scale=.3, baseline=-20pt]
    \coordinate (X1) at (-2, 0);
    \coordinate (X2) at (2, 0);
    \coordinate (Y1) at (0, 1);
    \coordinate (Y2) at (0, -1);
    \draw[line width=1.2pt] (0,0) ellipse (2 and 1);
    \fill (X1) circle (5pt);
    \fill (X2) circle (5pt);
   \node[font=\tiny] at (Y1) {$\times$};
    \node[font=\tiny] at (Y2) {$\times$};
    \draw[line width=1pt] (-2.4, 0)
        .. controls (-2.4, 1.35) and (-0.8, 1.35) .. (0.1, 1.35)  
        .. controls (0.6, 1.35) and (0.6, 0.65) .. (0.1, 0.65)    
        .. controls (-0.8, 0.65) and (-1.4, 0.65) .. (-1.4, 0)    
        .. controls (-1.4, -0.65) and (-0.8, -0.65) .. (0.1, -0.65) 
        .. controls (0.6, -0.65) and (0.6, -1.35) .. (0.1, -1.35) 
        .. controls (-0.8, -1.35) and (-2.4, -1.35) .. (-2.4, 0); 
    \node[left, font=\normalsize] at (-2.5, 0) {$X_1$};
    \node[right, font=\normalsize] at (2.2, 0) {$X_2$};
    \node[above, font=\normalsize] at (0, 1.5) {$Y_1$};
    \node[below, font=\normalsize] at (0, -1.5) {$Y_2$};
\end{tikzpicture}}}.
\end{align}
These precisely correspond to the tubes that are neither nested nor disjoint with $\Delta_{2,\bar3}$, which by the rules of kinematic flow cannot appear next to the corresponding letter in the symbol (this can also be checked explicitly from \eqref{eq: symbol bubble}). In other words, the property of cluster adjacency is implied by the kinematic flow rules, using our map \eqref{eq: map n-gon}.

We end by presenting here the full expression for the symbol~\eqref{eq: symbol bubble} expressed in terms of the chord variables~(\ref{eq:delta1}) and~(\ref{eq: cluster vars from boundary chords}):
\begin{multline}
\label{eq:s2loop-final}
\mathcal{S}_2^{(\text{loop})} =\Delta_{1,2}\otimes\Delta_{1,\bar2} - \Delta_{1,2}\otimes\Delta_{1,\bar3} - \Delta_{1,2}\otimes\Delta_{2,4} + \Delta_{1,2}\otimes\Delta_{3,4}
    - \Delta_{1,4}\otimes\Delta_{1,2} \\ + \Delta_{1,4}\otimes\Delta_{1,3}+ \Delta_{1,4}\otimes\Delta_{2,4} - \Delta_{1,4}\otimes\Delta_{3,4} - \Delta_{2,\bar3}\otimes\Delta_{1,2} + \Delta_{2,\bar3}\otimes\Delta_{1,\bar3}  + \Delta_{2,\bar3}\otimes\Delta_{2,\bar4} \\ - \Delta_{2,\bar3}\otimes\Delta_{3,4} + \Delta_{3,4}\otimes\Delta_{1,2} - \Delta_{3,4}\otimes\Delta_{1,3} - \Delta_{3,4}\otimes\Delta_{2,\bar4}+\Delta_{3,4}\otimes\Delta_{3,\bar4}\,,
\end{multline}
which manifestly satisfies cluster adjacency.

\subsection{Example: 3-site loop}
\label{sec:gon3}
For the wavefunction coefficient associated with the 3-site loop, or triangle graph, the symbol alphabet is associated with a type $B_5$ cluster algebra. It is composed of 28 letters (from which one can form 27 dimensionless ratios)  that correspond to the region variables one can draw as tubes in the graph, and is reproduced by the map:
\begin{align}
\begin{gathered}
    \Delta_{1,2}  = \frac{X_1{+}Y_1{+}Y_3}{X_1{+}X_2{+}X_3}\,,\quad \Delta_{2,3}= \frac{{-}2Y_1}{X_1{+}X_2{+}X_3}\,,\quad \Delta_{3,4} = \frac{X_2{+}Y_1{+}Y_2}{X_1{+}X_2{+}X_3}\,,\\ \Delta_{4,5}= \frac{{-}2Y_2}{X_1{+}X_2{+}X_3}\,,\quad \Delta_{5,6}  = \frac{X_3+Y_2+Y_3}{X_1+X_2+X_3}\,.
\end{gathered}
\end{align}
The remainder of the cluster variables are obtained using \eqref{eq: cluster vars from boundary chords} and $\Delta_{a,\bar b} = 1-\Delta_{a,b}$. Along with the complete alphabet, this also generates three spurious letters $-2Y_1$, $-2Y_2$ and $-2Y_3$.

As in the previous cases, this map also manifestly satisfies cluster adjacency as a direct consequence of the rules of kinematic flow. To illustrate this, consider the cluster variable:
\begin{equation}
    \Delta_{2,4} = \frac{X_2{-}Y_1{+}Y_2}{X_1{+}X_2{+}X_3} = \vcenter{\hbox{\begin{tikzpicture}[
    scale=0.4, 
    baseline={(current bounding box.center)}, 
    every node/.style={font=\small}
]

    \coordinate (X1) at (-1.5, 0);
    \coordinate (X2) at (0, 2.6);
    \coordinate (X3) at (1.5, 0);
    
    \coordinate (Y1) at (-0.75, 1.3);
    \coordinate (Y2) at (0.75, 1.3);
    \coordinate (Y3) at (0, 0);

    \draw[line width=1.2pt] (X1) -- (X2) -- (X3) -- cycle;

    \fill (X1) circle (5pt);
    \fill (X2) circle (5pt);
    \fill (X3) circle (5pt);

    \node[font=\small] at (Y1) {$+$};
    \node[font=\small] at (Y2) {$+$};
    \node[font=\small] at (Y3) {$\times$};

    \draw[line width=1pt] plot [smooth cycle, tension=0.6] coordinates {
        (-1.0, 0.9)   
        (-1.15, 1.4)  
        (-0.4, 2.7)   
        (0.05, 2.95)  
        (0.35, 2.5)   
        (-0.3, 1.1)   
    };

    \node[below left] at (X1) {$X_1$};
    \node[above, yshift=5pt] at (X2) {$X_2$};
    \node[below right] at (X3) {$X_3$};
    
    \node[left, xshift=-5pt, yshift=5pt] at (Y1) {$Y_1$};
    \node[right, xshift=5pt, yshift=5pt] at (Y2) {$Y_2$};
    \node[below, yshift=-5pt] at (Y3) {$Y_3$};

\end{tikzpicture}}}.
\end{equation}
Geometrically, it can be represented as the pair of chords $(2,4)$ and $(\bar2,\bar4)$ in a dodecagon with vertices labeled $\{1,\ldots,6,\bar1,\ldots,\bar6\}$. The incompatible cluster variables then correspond to all pairs of chords that cross this one. For example, two such variables are:
\begin{equation}
    \begin{aligned}
        \Delta_{3,\bar5} = \frac{X_1{+}X_3{-}Y_1{+}Y_2}{X_1{+}X_2{+}X_3} = \vcenter{\hbox{\begin{tikzpicture}[
    scale=0.4, 
    baseline={(current bounding box.center)}, 
    every node/.style={font=\small}
]

    \coordinate (X1) at (-1.5, 0);
    \coordinate (X2) at (0, 2.6);
    \coordinate (X3) at (1.5, 0);
    
    \coordinate (Y1) at (-0.75, 1.3);
    \coordinate (Y2) at (0.75, 1.3);
    \coordinate (Y3) at (0, 0);

    \draw[line width=1.2pt] (X1) -- (X2) -- (X3) -- cycle;

    \fill (X1) circle (5pt);
    \fill (X2) circle (5pt);
    \fill (X3) circle (5pt);

    \node[font=\small] at (Y1) {$+$};
    \node[font=\small] at (Y2) {$+$};
    \node[font=\small] at (Y3) {$\times$};

    \draw[line width=1pt] plot [smooth cycle, tension=0.6] coordinates {
        (-0.85, 1.75) 
        (-1.65, 0.6)   
        (-1.9, -0.4)  
        (0, -0.45)     
        (1.65, -0.4)  
        (1.85, 0.1)    
        (1.3, 0.45)    
        (0, 0.4)      
        (-0.75, 0.4) 
        (-0.45, 1.1)   
        (-0.35, 1.6)   
    };

    \node[below left, xshift=-2pt, yshift=-2pt] at (X1) {$X_1$};
    \node[above, yshift=5pt] at (X2) {$X_2$};
    \node[below right, xshift=2pt, yshift=-2pt] at (X3) {$X_3$};
    
    \node[left, xshift=-5pt, yshift=5pt] at (Y1) {$Y_1$};
    \node[right, xshift=5pt, yshift=5pt] at (Y2) {$Y_2$};
    \node[below, yshift=-7pt] at (Y3) {$Y_3$};

\end{tikzpicture}}},\ \ 
\Delta_{3,6} = \frac{X_2{+}X_3{+}Y_1{+}Y_3}{X_1{+}X_2{+}X_3} = \vcenter{\hbox{\begin{tikzpicture}[
    scale=0.4, 
    baseline={(current bounding box.center)}, 
    every node/.style={font=\small}
]

    \coordinate (X1) at (-1.5, 0);
    \coordinate (X2) at (0, 2.6);
    \coordinate (X3) at (1.5, 0);
    
    \coordinate (Y1) at (-0.75, 1.3);
    \coordinate (Y2) at (0.75, 1.3);
    \coordinate (Y3) at (0, 0);

    \draw[line width=1.2pt] (X1) -- (X2) -- (X3) -- cycle;

    \fill (X1) circle (5pt);
    \fill (X2) circle (5pt);
    \fill (X3) circle (5pt);

    \node[font=\small] at (Y1) {$+$};
    \node[font=\small] at (Y2) {$+$};
    \node[font=\small] at (Y3) {$\times$};

    \draw[line width=1pt] plot [smooth cycle, tension=0.6] coordinates {
        (-0.4, 2.6)   
        (0.05, 3)  
        (0.4, 2.7)    
        (1.15, 1.4)   
        (1.85, -0.1)   
        (1.4, -0.4)  
        (1.05, -0.05) 
        (0.3, 1.25)   
    };

    \node[below left] at (X1) {$X_1$};
    \node[above, yshift=5pt] at (X2) {$X_2$};
    \node[below right, xshift=2pt, yshift=-1.5pt] at (X3) {$X_3$};
    
    \node[left, xshift=-5pt, yshift=5pt] at (Y1) {$Y_1$};
    \node[right, xshift=5pt, yshift=5pt] at (Y2) {$Y_2$};
    \node[below, yshift=-5pt] at (Y3) {$Y_3$};

\end{tikzpicture}}}.
    \end{aligned}
\end{equation}
Clearly, these are associated with tubes that are neither nested nor disjoint with the first one, which is consistent with kinematic flow. We have explicitly checked that the complete symbol for the triangle graph
indeed satisfies cluster adjacency.

\subsection{Bootstrap}
\label{sec:gonboot}

We consider the cluster bootstrap for the symbol of the $n$-site loop cosmological wavefunction coefficient. The bootstrap constraints are similar to those used above for the $n$-site chain; the only difference is that the discrete
symmetry of the symbol is not the flip symmetry~(\ref{eq:flip}) but the full dihedral group acting on labels of the $n$-gon, generated by:
\begin{equation}
    \text{flip: } X_i\leftrightarrow X_{n-i+1}\,,\ Y_{i} \leftrightarrow Y_{n-i}\,,\qquad \text{cyclic: } X_i \to X_{i+1}\,,\ Y_{i}\to Y_{i+1}\,,
\end{equation}
with all indices understood mod $n$ as usual.

Let us describe the bootstrap for the bubble in detail. In this case, including the spurious letters, there are 12 dimensionless letters:
\begin{equation}
\begin{aligned}
    \Phi_{1,\dots,12}= \bigg\{ & \frac{X_1+Y_1+Y_2}{X_1+X_2},\, \frac{X_1-Y_1+Y_2}{X_1+X_2},\, \frac{X_1+X_2+2 Y_2}{X_1+X_2},\, \frac{X_2-Y_1-Y_2}{X_1+X_2}, \\
    &\frac{X_2+Y_1-Y_2}{X_1+X_2},\, \frac{-2 Y_2}{X_1+X_2},\, \frac{-2 Y_1}{X_1+X_2},\, \frac{X_2-Y_1+Y_2}{X_1+X_2},\\
    & \frac{X_1+X_2+2 Y_1}{X_1+X_2},\, \frac{X_1+Y_1-Y_2}{X_1+X_2},\, \frac{X_2+Y_1+Y_2}{X_1+X_2},\, \frac{X_1-Y_1-Y_2}{X_1+X_2}\bigg\} \\ 
    = \{ & \Delta_{1,2},\, \Delta_{1,3},\, \Delta_{1,4},\, \Delta_{1,\bar2},\, \Delta_{1,\bar3},\, \Delta_{1,\bar4},\, \Delta_{2,3},\, \Delta_{2,4},\, \Delta_{2,\bar3},\, \Delta_{2,\bar4},\, \Delta_{3,4},\, \Delta_{3,\bar4}\}\,.
\end{aligned}
\end{equation}
We find that the space of integrable weight-2 symbols in this alphabet is 96-dimensional and the subspace that obeys the first-entry condition (that each term in the symbol must begin with $\Phi_1$, $\Phi_3$, $\Phi_9$ or $\Phi_{11}$) is 18-dimensional. We further reduce the space by imposing dihedral symmetry and the soft limit, which leaves a 2-dimensional space spanned by 
\begin{align}
\Delta_{1,4}\otimes\Delta_{2,\bar3}+\Delta_{2,\bar3}\otimes\Delta_{1,4}\,,
\end{align}
which is ruled out by cluster adjacency, and by the correct answer~(\ref{eq:s2loop-final}).
Thus, the bootstrap uniquely determines the correct result for the
symbol (up to overall normalization).

\begin{table}
\begin{longtable}{|C{5cm}|C{1cm}|C{1cm}|C{1cm}|C{1cm}|C{1cm}|C{1cm}|}
    \hline
    Constraints
    & \multicolumn{2}{c|}{2-site loop}
    & \multicolumn{2}{c|}{3-site loop}
    & \multicolumn{2}{c|}{4-site loop} \\
    \hline
    Integrability and first entry & 18 & 18 & 546 & 546 & 23780 & 23780 \\
    \hline
    Adjacency & 17 & N/I & 379 & N/I & 9601 & N/I\\
    \hline
    Dihedral Symmetry & 7 & 8 & 76 & 111 & 1267 & 3137\\
    \hline
    $Y_i \to 0$ limit & 1 & 2 & 1 & 3 & 1 & 7 \\
    \hline
    \caption{A summary of the bootstrap for the 2-, 3- and 4- site loop graphs. The symbols have transcendental weight 2, 3 and 4, respectively and the initial alphabets have 12, 30, and 56 dimensionless letters. There are two columns for each case; the answer is uniquely determined (up to an overall constant) with cluster adjacency imposed (left columns) but not fixed when it is omitted (right columns). In all three cases the spurious letters drop out automatically after imposing the final $Y_i \to 0$ constraint.}
    \label{tab: loop summary}
\end{longtable}
\end{table}

We have carried out the $n$-gon bootstrap for $n=2, 3, 4$.  In all three cases, we have found
there is a unique (up to overall normalization) symbol that satisfies all of the bootstrap constraints
1--6 reviewed in Section~\ref{sec:chainboot}, the spurious letter(s) drop out automatically (constraint 7), and this unique answer agrees with the correct answer for the symbol of the corresponding de Sitter wavefunction coefficient.
In contrast to the case of the $n$-site chain described there, for the $n$-gon we find that imposing cluster adjacency is necessary in order to obtain a unique answer. This is evidenced in the last rows of Table~\ref{tab: loop summary}, where we summarize the dimensions of the symbol spaces for this cluster bootstrap.

Finally, we summarize the counting of ordered adjacent pairs of letters allowed by cluster compatibility and the number of ordered pairs that appear in the symbol of polygonal graphs in Table \ref{tab: loop compatible counting}.
\begin{table}[h!]
    \centering
    \begin{tabular}{|c|c|c|}
        \hline
        Example & Cluster compatible pairs & Wavefunction letter pairs \\
        \hline
        2-site loop & 108 & 16\\
        \hline
        3-site loop & 600 & 168\\
        \hline
        4-site loop & 1960 & 640\\
        \hline
    \end{tabular}
    \caption{A counting of cluster compatible letter pairs and pairs that appear in the polygon wavefunction coefficient.}
    \label{tab: loop compatible counting}
\end{table}

\section{Outlook}
\label{sec:outlook}

In this paper we have shown that the symbol alphabets for $n$-site chain and loop graphs contributing to de Sitter wavefunction coefficients for a cubic scalar field are subsets of the cluster variables of the $A_{2{n}-2}$ and $B_{2n{-}1}$ cluster algebras, respectively. This observation was recently made for the chain graphs in~\cite{Capuano:2025myy,Mazloumi:2025pmx}. The map allows us to further prove that the symbols associated with both chain and loop graphs satisfy cluster adjacency with respect to the corresponding $A$- and $B$-type algebras. The key step in the proof is identifying a precise correspondence between complete tubings, that appear in describing the kinematic flow equation~(\ref{eq:dAI}), and clusters: the former are maximal sets of non-overlapping tubes on a graph, and the latter are maximal sets of non-overlapping chords in a triangulation of a polygon. In this connection each individual tube in a complete tubing corresponds to a chord in the triangulation, and the kinematic flow equations guarantee that the symbol letters corresponding to two different tubes can only appear next to each other in a symbol if they are nested or disjoint, which is equivalent to the condition that the corresponding chords do not intersect, and agrees with the mathematical notion of cluster compatibility.

Finally, we also discussed the use of a bootstrap to compute the symbols of these wavefunction coefficients by combining these cluster properties with basic physical input including the first-entry condition, discrete symmetries, and vanishing in the soft limit.  We find that these properties uniquely determine the symbols of all $n$-site chain and loop graphs for $n = 2, 3, 4$ up to overall coefficients. We expect that these overall coefficients will be fixed by demanding that the symbol satisfy cosmological cutting rules such as those given in~\cite{Baumann:2021fxj, Melville:2021lst}.

It would be interesting to generalize our results in a number of ways.  First, it would be interesting to see which cluster algebras (if any) more general graphs correspond to.  Second, our results apply only to massless fields, but it is useful to note that progress has been made in the study of differential equations for cosmological wavefunction coefficients with massive states~\cite{Gasparotto:2024bku}. Indeed in a low mass expansion, it was found that each order only contains multiple polylogarithm functions, which suggests that the study of the symbol and corresponding cluster structure might be fruitful.

Let us comment on the applicability of our results beyond de Sitter space. The differential equation as we have written it in~(\ref{eq:dAI}) is only the $\epsilon \to 0$ truncation of the full kinematic flow equation~\cite{Arkani-Hamed:2023kig} that applies to wavefunction coefficients in general FRW backgrounds with cosmological scale factor $(\eta/\eta_0)^{-1 + \epsilon}$.  In general, the solution to the differential equation can be written as an expansion in $\epsilon$ where each term is a multiple polylogarithm function (of increasing weight at higher orders in $\epsilon$).
The set of symbol letters and the structure of the differential equations, in particular regarding which symbol letters can appear adjacent to which others, are encoded in $A$ and remain the same for general $\epsilon$. Thus, we expect that one should encounter only cluster adjacent polylogarithm functions to all orders in the $\epsilon$ expansion around de Sitter space\footnote{This is only true if the interaction vertex is cubic.  For a scalar $\phi^p$ interaction it can be checked that the analogous expansion is not around de Sitter ($\epsilon=0$), but rather the perhaps less physically significant background where $\epsilon=(p-3)/(4-p)$.}. Further, the bootstrap conditions 1--7 tabulated in Sections~\ref{sec:chainboot} and~\ref{sec:gonboot} also remain unchanged, and it would be interesting to investigate the systematic bootstrap of these terms in the $\epsilon$ expansion.

In the context of $\mathcal{N}=4$ SYM theory it has been a longstanding problem to understand the precise connection between 
the cluster algebra structure of (integrated) multi-loop amplitudes and that of their integrands.  In the context of cosmology,
the flat-space wavefunction coefficients serve as
the ``integrands'' for those of
de Sitter space (or more general FRW cosmologies)~\cite{Arkani-Hamed:2017fdk}.
It can be seen from the polytope constructions of the $n$-site chains and $n$-gons in flat space~\cite{De:2025bmf} that the number of boundaries corresponds to the number of cluster variables in the $A_{n{-}2}$ and $B_{n{-}1}$ cluster algebras, respectively. In this paper we have demonstrated that
the cosmological wavefunction coefficients, obtained from these flat space coefficients
after integration, have symbol alphabets that are respectively subsets of the $A_{2n{-}2}$ and $B_{2n{-}1}$ cluster algebras.
It would be fascinating to see if this is an arena where the understanding of the cluster structure of integrands
and integrals can be made more precise.

\color{black}

\acknowledgments

We are grateful to Shounak De and Andrzej Pokraka for useful discussions. This work was supported in part by the US Department of Energy under contract DE-SC0010010 Task F and by Simons Investigator Award \#376208 (SP, AV).

\bibliographystyle{JHEP}
\bibliography{biblio.bib}

\end{document}